\documentclass[a4paper,12pt]{article}
\usepackage{amssymb}
\usepackage{amsmath}
\usepackage{latexsym}
\usepackage[dvips]{graphicx}
\usepackage{cite}
\newcommand{\be}{\begin{equation}}
\newcommand{\ee}{\end{equation}}
\newcommand{\bea}{\begin{eqnarray}}
\newcommand{\nn}{\nonumber}
\newcommand{\eea}{\end{eqnarray}}

\begin{document}

\begin{titlepage}

\begin{centering}
\vspace{.3in}
{\Large{\bf Energy and Momentum Distributions \\ of the Magnetic
Solution to (${\bf 2+1}$) Einstein-Maxwell Gravity}}
\\

\vspace{.5in} {\bf  Th. Grammenos\footnote{thgramme@uth.gr} }\\

\vspace{0.3in}
Laboratory of Fluid Mechanics \& Turbomachines\\
Dept. of Mechanical and Industrial Engineering\\
University of Thessaly\\
38334 Volos,
Greece\\
\end{centering}

\vspace{0.7in}
%%%%%%%%%%%%%%%%%%%ABSTRACT%%%%%%%%%%%%%%%%%%%%%%%%%%%%%%%%%%
\begin{abstract}
\par\noindent
We use M{\o}ller's energy-momentum complex in order to explicitly
evaluate the energy and momentum density distributions associated
with the three-dimensional magnetic solution to the
Einstein-Maxwell equations. The magnetic spacetime under
consideration is a one-parametric solution describing the
distribution of a radial magnetic field in a three-dimensional AdS
background, and representing the superposition of the magnetic
field with a 2+1 Einstein static gravitational field.
\end{abstract}

%%%%%%%%%%%%%%%%%%%%%%%%%%%%%%%%%%%%%%%%
\end{titlepage}
\newpage

\baselineskip=18pt
%%%%%%%%%%%%%%%%%%%%%%%%%%%%%%%%%%%%%%%%%%%%%%%%%%%%%%%%%%%%%%%%%%%%%%%%%%%%%%%%%%%%%%%%%%%%%%%%%
%%%%%%%%%%%%%%%%%%%%%%%%%%%%%%%%%%%%%%%%%%%%%%%%%%%%%%%%%%%%%%%%%%%%%%%%%%%%%%%%%%%%%%%%%%%%%%%%
%%%%%%%%%%%%%%%%%%%%%%%%%%%%%%%%%%%%%%%%%%%%%%%%%%%%%%%%%%%%%%%%%%%%%%%%%%%%%%%%%%%%%%%%%%%%%%%%%
%%%%%%%%%%%%%%%%%%%%%%%%%%% INTRODUCTION
%%%%%%%%%%%%%%%%%%%%%%%%%%%%%%%%%%%%%%%%%%%%%%%%%%%%%%%%
\section*{Introduction}
One of the oldest problems in gravitation which still lacks of a
definite answer is the localization of energy and momentum. Much
attention has been devoted for this problematic issue. Einstein
was the first to construct a locally conserved energy-momentum
complex \cite{einstein}. Consequently, a plethora of different
energy-momentum complexes  were proposed
\cite{ll}-\hspace{-0.1ex}\cite{weinberg}. These expressions were
restricted to evaluate energy distribution in quasi-Cartesian
coordinates. M{\o}ller \cite{moller} proposed a new expression for
an energy-momentum complex which could be utilized to any
coordinate system. However, the idea of the energy-momentum
complex was severely criticized for a number of reasons. Firstly,
although a symmetric and locally conserved object, its nature is
nontensorial and thus its physical interpretation seemed obscure
\cite{chandra}. Secondly, different energy-momentum complexes
could yield different energy distributions for the same
gravitational background \cite{bergqvist1,bergqvist2}. Thirdly,
energy-momentum complexes were local objects while there was
commonly believed that the proper energy-momentum of the
gravitational field was only total, i.e. it cannot be localized
\cite{chiang}. For a long time, attempts to deal with this
problematic issue were made only by proposers of quasi-local
approach \cite{brown,sean}.

\par\noindent
In 1990 Virbhadra revived the interest in this approach
\cite{virbhadra}. At the same time Bondi \cite{bondi} sustained
that a nonlocalizable form of energy is not admissible in
relativity so its location can in principle be found. Since then,
numerous works on evaluating the energy distribution of several
gravitational backgrounds have been completed employing the
abandoned for a long time approach of energy-momentum complexes
\cite{complexes}.

%\par\noindent
In 1996 Aguirregabiria, Chamorro and Virbhadra \cite{virbhadra1}
showed that five different\footnote{Later on Virbhadra
\cite{virbhadra2} came  to know that Tolman's and Einstein's
complexes which had been used in \cite{virbhadra1} were exactly
the same (see footnote 1 in \cite{virbhadra2}).} energy-momentum
complexes yield the same energy distribution for any Kerr-Schild
class metric. Additionally, their results were identical with the
results of Penrose \cite{pen} and Tod \cite{tod} using the notion
of quasi-local mass.

%\par\noindent
Later attempts to deal with this problematic issue were made (as
already mentioned) by proposers of quasi-local approach. The
determination as well as the computation of the quasilocal energy
and quasilocal angular momentum of a ($2+1$)-dimensional
gravitational background were first presented by Brown, Creighton
and Mann \cite{mann1}. Many attempts since then have been
performed to give new definitions of quasilocal energy in General
Relativity \cite{quasilocal}. Considerable efforts  have also been
performed in constructing superenergy tensors \cite{senovilla1}.

%\par\noindent
In 1999 Chang, Nester and Chen \cite{nester} proved that every
energy-momentum complex is associated with a Hamiltonian boundary
term. Thus, the energy-momentum complexes are quasi-local and
acceptable.

%\par\noindent
In this work the approach of energy-momentum complexes is
implemented. The gravitational background  under investigation is
the ($2+1$)-dimensional rotating magnetic spacetime \cite{catal}
of the Einstein-Maxwell gravity. We evaluate the energy confined
in a ``one-sphere'' ($S^{1}$) of radius $r_0$ associated with the
aforesaid background. Specifically, we are implementing the
prescription of M{\o}ller. The specific ($2+1$)-dimensional spacetime background is described by one self-consistent integration
constant, $\tilde{q}_m$. When $\tilde{q}_m =0$, the Anti-de
Sitter space is obtained. Additionally, the corresponding metric is horizonless and has no curvature singularity at the
origin.
%\par\noindent
The rest of the paper is organized as follows. In the first two
sections we consider the concept of energy-momentum complexes in
the context of General Theory of Relativity and give M{\o}ller's
prescription for the energy-momentum complex. In Section 3 we
briefly present the $(2+1)$-dimensional BTZ and AdS black holes, and
we give the magnetic solution to the 2+1
Einstein-Maxwell gravity, while in Section 4, using M{\o}ller's
energy-momentum complexes, we explicitly compute the energy and
momentum distributions contained in a ``one-sphere" of fixed
radius $r_0$, as well as the effective gravitational mass of the
spacetime under study. Additionally, the energy of AdS$_3$
spacetime is evaluated using M{\o}ller's complex and the
result is identical with that obtained by setting $\tilde{q}_m =0$
in the expression for the energy associated with the magnetic solution
to the 2+1 Einstein-Maxwell gravity. Finally, in Section 5 a
summary of the obtained results and some concluding remarks are presented.
%%%%%%%%%%%%%%%%%%%%%%%%%%%%%%%%%%%%%%%%%%%%%%%%%%%%%%%%%%%%%%%%%%%%%%%%%%%%%%%%%%%%%%%%%%%%%%%%%
%%%%%%%%%%%%%%%%%%%%%%%%%%%%%%%%%%%%%%%%%%%%%%%%%%%%%%%%%%%%%%%%%%%%%%%%%%%%%%%%%%%%%%%%%%%%%%%%%
%%%%%%%%%%%%%%%%%%%%%%%%%%%%%%%%%%%%%%%%%%%%%%%%%%%%%%%%%%%%%%%%%%%%%%%%%%%%%%%%%%%%%%%%%%%%%%%%%
\section{Energy-Momentum Complexes}
The conservation laws of energy and momentum  for an isolated,
i.e., no external force acting on the system, physical system in
the Special Theory of Relativity are expressed by a set of
differential equations. Defining $T^{\mu}_{\nu}$ as the symmetric
energy-momentum tensor of matter and of all non-gravitational
fields, the conservation laws are given by \be\label{1}
T^{\mu}_{\nu,\, \mu} \equiv \frac{\partial T^{\mu}_{\nu}}{\partial
x^{\mu}}=0 \ee where \be\label{2}
\rho=T^{t}_{t}\hspace{1cm}j^{i}=T^{i}_{t}\hspace{1cm}p_{i}=-T^{t}_{i}
\ee are the energy density, the energy current density, and the
momentum density, respectively, and Greek indices run over the
spacetime labels while Latin indices run over the spatial
coordinate values.
%\par\noindent
Making the transition from Special to General Theory of
Relativity, one adopts a simplicity principle which is called
principle of minimal gravitational coupling. As a result of this,
the conservation equation is now written as \be\label{3}
T^{\mu}_{\nu;\, \mu} \equiv
\frac{1}{\sqrt{-g}}\frac{\partial}{\partial
x^{\mu}}\left(\sqrt{-g}\,T^{\mu}_{\nu}\right)-\Gamma^{\kappa}_{\nu\lambda}T^{\lambda}_{\kappa}=0
\ee where $g$ is the determinant of the metric tensor
$g_{\mu\nu}(x)$. The conservation equation may also be written as
\be\label{4} \frac{\partial}{\partial
x^{\mu}}\left(\sqrt{-g}\,T^{\mu}_{\nu}\right)=\xi_{\nu} \ee where
\be\label{5}
\xi_{\nu}=\sqrt{-g}\Gamma^{\kappa}_{\nu\lambda}T^{\lambda}_{\kappa}
\ee is a non-tensorial object. For $\nu=t$ this means that the
matter energy is not a conserved quantity for the physical
system\footnote{It is possible to restore the conservation law by
introducing a local inertial system for which at a specific
spacetime point $\xi_{\nu}=0$ but this equality by no means holds
in general.}. From a physical point of view, this lack of energy
conservation can be understood as the possibility of transforming
matter energy into gravitational energy and vice versa. However,
this remains an open problem and it is widely believed that in
order to solve it one has to take into account the gravitational
energy.

%\par\noindent
By a well-known procedure, the non-tensorial object $\xi_{\nu}$
can be written  as \be\label{6}
\xi_{\nu}=-\frac{\partial}{\partial
x^{\mu}}\left(\sqrt{-g}\,\vartheta^{\mu}_{\nu}\right) \ee where
$\vartheta^{\mu}_{\nu}$ are certain functions of the metric tensor
and its first order derivatives. Therefore, the energy-momentum
tensor of matter $T^{\mu}_{\nu}$ is replaced by the expression
\be\label{7}
\theta^{\mu}_{\nu}=\sqrt{-g}\left(T^{\mu}_{\nu}+\vartheta^{\mu}_{\nu}\right)
\ee which is called energy-momentum complex since it is a
combination of the tensor $T^{\mu}_{\nu}$ and a pseudotensor
$\vartheta^{\mu}_{\nu}$ describing the energy and  momentum of the
gravitational field. The energy-momentum complex satisfies a
conservation law in the ordinary sense, i.e., \be\label{8}
\theta^{\mu}_{\nu,\, \mu}=0 \ee and it can be written as
\be\label{9}
\theta^{\mu}_{\nu}=\chi^{\mu\lambda}_{\nu\,\,\,,\lambda} \ee where
$\chi^{\mu\lambda}_{\nu}$ are called superpotentials and are
functions of the metric tensor and its first order derivatives.

%\par\noindent
It is evident that the energy-momentum complex is not uniquely
determined by the condition that its usual divergence is zero
since a quantity with an identically vanishing divergence can
always be added to the energy-momentum complex.
%%%%%%%%%%%%%%%%%%%%%%%%%%%%%%%%%%%%%%%%%%%%%%%%%%%%%%%%%%%%%%%%%%%%%%%%%%%%%%%%%%%%%%%%%%%%%%%%%
%%%%%%%%%%%%%%%%%%%%%%%%%%%%%%%%%%%%%%%%%%%%%%%%%%%%%%%%%%%%%%%%%%%%%%%%%%%%%%%%%%%%%%%%%%%%%%%%%
%%%%%%%%%%%%%%%%%%%%%%%%%%%%%%%%%%%%%%%%%%%%%%%%%%%%%%%%%%%%%%%%%%%%%%%%%%%%%%%%%%%%%%%%%%%%%%%%%
%%%%%%%%%%%%%%%%%%%%%%%%%%%%%%%%%%%%%%%%%%%%%%%%%%%%%%%%%%%%%%%%%%%%%%%%%%%%%%%%%%%%%%%%%%%%%%%%%
\section{M{\o}ller's Prescription}
The energy-momentum complex of M{\o}ller in a  four-dimensional
background is given as \cite{moller} \be\label{mtheta}
\mathcal{J}^{\mu}_{\nu}=\frac{1}{8\pi}\xi^{\mu\lambda}_{\nu\,\,
,\, \lambda}  \ee where the M{\o}ller's superpotential $
\xi^{\mu\lambda}_{\nu}$ is of the form \be\label{msuper}
\xi^{\mu\lambda}_{\nu}=\sqrt{-g} \left(\frac{\partial
g_{\nu\sigma} }{\partial x^{\kappa} }- \frac{\partial
g_{\nu\kappa}}{\partial x^{\sigma}
}\right)g^{\mu\kappa}g^{\lambda\sigma}  \ee with the antisymmetric
property \be\label{12}
\xi^{\mu\lambda}_{\nu}=-\xi^{\lambda\mu}_{\nu}\hspace{1ex}. \ee

%\par\noindent
It is easily seen that the M{\o}ller's energy-momentum complex
satisfies the local conservation equation \be\label{13}
\frac{\partial \mathcal{J}^{\mu}_{\nu}}{\partial x^{\mu}}=0 \ee
where $\mathcal{J}^{0}_{0}$ is the energy density and
$\mathcal{J}^{0}_{i}$ are the momentum density components.

%\par\noindent
Thus, in M{\o}ller's prescription the energy and momentum for a
four-dimensional background are given by \be\label{mmomentum}
P_{\mu}=\int\int\int \mathcal{J}^{0}_{\mu}dx^{1}dx^{2}dx^{3}  \ee
and specifically the energy of the physical system in a
four-dimensional background is \be\label{menergy} E=\int\int\int
\mathcal{J}^{0}_{0}dx^{1}dx^{2}dx^{3}\hspace{1ex}. \ee

%\par\noindent
It should be noted that the calculations are not anymore
restricted to quasi-Cartesian coordinates but can be utilized in
any coordinate system.
%%%%%%%%%%%%%%%%%%%%%%%%%%%%%%%%%%%%%%%%%%%%%%%%%%%%%%%%%%%%%%%%%%%%%%%%%%%%%%%%%%%%%%%%%%%%%%%%%
%%%%%%%%%%%%%%%%%%%%%%%%%%%%%%%%%%%%%%%%%%%%%%%%%%%%%%%%%%%%%%%%%%%%%%%%%%%%%%%%%%%%%%%%%%%%%%%%%
%%%%%%%%%%%%%%%%%%%%%%%%%%%%%%%%%%%%%%%%%%%%%%%%%%%%%%%%%%%%%%%%%%%%%%%%%%%%%%%%%%%%%%%%%%%%%%%%%
%%%%%%%%%%%%%%%%%%%%%%%%%%%%%%%%%%%%%%%%%%%%%%%%%%%%%%%%%%%%%%%%%%%%%%%%%%%%%%%%%%%%%%%%%%%%%%%%%
%%%%%%%%%%%%%%%%%%%%%%%%%%%%%%%%%%%%%%%%%%%%%%%%%%%%%%%%%%%%%%%%%%%%%%%%%%%%%%%%%%%%%%%%%%%%%%%%%
\section{AdS$_3$ Black Holes and the Magnetic Solution to the (${\bf 2+1}$) Einstein-Maxwell Gravity}
In 1992 Ba\~{n}ados, Teitelboim, and Zanelli discovered a black
hole solution (known as  BTZ black hole) in ($2+1$) dimensions
\cite{jorge}. Till that time it was believed that no black hole
solution exists in three-dimensional spacetimes \cite{abbott}.
Ba\~{n}ados, Teitelboim, and Zanelli found a vacuum solution to
Einstein gravity with a negative cosmological constant.

%\par\noindent
The starting point was the action in a three-dimensional theory of
gravity \be\label{action} S=\int
d^{3}x\sqrt{-g}\left(R+\frac{2}{l^{2}}\right) \ee where the radius
of curvature $l$ is related to the cosmological constant by
$\Lambda=-l^{-2}$.

%\par\noindent
It is straightforward to check that Einstein's field equations
\be\label{fieldeqns} R_{\mu\nu}-\frac{1}{2}g_{\mu\nu}\left(R+
\frac{2}{l^{2}}\right) = 0 \ee are solved by the
metric\footnote{The form of the BTZ metric in quasi-Cartesian
coordinates can be obtained by making the transformations \bea
x&=&r\cos (\phi)\nn\\
y&=&r\sin(\phi)\nn \hspace{1ex}. \eea} \be\label{metric}
ds^{2}=N^{2}(r)dt^{2}
-\frac{dr^{2}}{N^{2}(r)}-r^{2}\left(N^{\phi}(r)dt+d\phi\right)^{2},
\ee where the squared lapse $N^{2}(r)$ and the angular shift
$N^{\phi}(r)$ are given by \be \label{lapse}
N^{2}(r)=\frac{r^2}{l^2}- M
+\frac{J^{2}}{4r^{2}}\hspace{1ex},\hspace{4ex}
N^{\phi}(r)=-\frac{J}{2r^{2}}  \ee with $-\infty<t<+\infty$,
$0<r<+\infty$, and $0\leq\phi\leq 2\pi$. Since the metric
(\ref{metric}) satisfies Einstein's field equations with a
negative cosmological constant (see (\ref{fieldeqns})), the metric
is locally Anti-de Sitter \be\label{20} ds^{2}=\left(1+
\frac{r^2}{l^2}\right)dt^{2}-\frac{dr^{2}}{\left(1+\displaystyle{\frac{r^2}{l^2}}\right)}
-r^{2}d\phi ^{2} \ee and it can only differ from Anti-de Sitter
space in its global properties. The two constants $M$ and $J$ are
the conserved quantities mass and angular momentum, respectively.
The lapse function $N(r)$ vanishes for two values of the radial
coordinate $r$ given by \be\label{21}
r^{2}_{\pm}=\frac{l^2}{2}\left[M\pm \sqrt{M^2 -
\left(\frac{J}{l}\right)^2}\right] \hspace{1ex}. \ee The largest
root, $r_{+}$, gives the black hole horizon. It is evident that in
order for the horizon to exist one must have \be\label{22} M>0
\hspace{1ex},\qquad \qquad |J|\leq M l \hspace{1ex}. \ee Therefore, negative
black hole masses are excluded from the physical spectrum. There
is, however, an important exceptional case. When one sets $M=-1$
and $J=0$, the singularity, i.e., $r=0$, disappears. There is
neither a horizon nor a singularity to hide. The configuration is
again that of Anti-de Sitter space. Thus, Anti-de Sitter emerges
as a ``bound state'', separated from the continuous black hole
spectrum by a mass gap of one unit.
%\par\noindent
For the specific case of spinless ($J=0$) BTZ black hole, the line
element (\ref{metric}) takes the simple form \be\label{spinless}
ds^{2}=\left(\frac{r^{2}}{l^{2}}- M\right)dt^{2}
-\frac{dr^{2}}{\left(\displaystyle{\frac{r^{2}}{l^{2}}}- M\right)}
-r^{2}d\phi^{2} \hspace{1ex}.  \ee As it is stated in the
Introduction, metric (\ref{metric}) of the rotating
($2+1$)-dimensional BTZ black hole is not asymptotically (that is
as $r\rightarrow \infty$) flat \be\label{24} ds^{2}=dt^2 -dr^2
-d\phi^{2} \hspace{1ex}. \ee

The (2+1)-dimensional magnetic solution to the
Einstein-Maxwell field equations has been given first by Clement \cite{clement}, followed by Peld\'{a}n \cite{peldan}, Hirschmann
and Welch \cite{hirschmann} and Cataldo and Salgado \cite{cataldo}. In the present work, we use the generalisation
to the rotating case as formulated by Dias and Lemos
\cite{dias}, whereby the line element is in the form
\begin{equation}\label{28}
\begin{split}
ds^2 &= \left(\frac{r^2}{l^2}-M\right)dt^2\\
\\
&-\frac{r^2}{(\displaystyle{\frac{r^2}{l^2}-M})(r^2+Q^2_m
\ln|\displaystyle{\frac{r^2}{l^2}-M}|)}\, dr^2\\
\\
&- \left(r^2+Q^2_m \ln|\displaystyle{\frac{r^2}{l^2}-M}|\right)\,
d\phi^2
\end{split}
\end{equation}
with $l=-1/\sqrt{\Lambda}$,  the radius of a pseudo-sphere,
$\Lambda$ the cosmological constant, and $Q_m$, $M$
self-consistent integration constants of the Einstein-Maxwell
field equations. In the case $Q_m=0$, the metric (\ref{28})
reduces to the spinless three-dimensional BTZ-black hole
(\ref{spinless}). However, Cataldo et al \cite{catal} have shown
recently, that the field parameter related to the mass of the
solution (\ref{28}) is a pure gauge and can be rescaled to -1.
Thus, the magnetic metric describing a distribution of a radial
magnetic field in an AdS$_3$ spacetime is given by the line
element
\begin{equation}\label{29}
ds^2 =\left(\frac{r^{\prime \, 2}}{l^2}+1\right)\, dt^{\prime\, 2}
-\frac{r^{\prime\, 2}}{(\displaystyle{\frac{r^{\prime\,
2}}{l^2}}+1)F^{\prime \, 2}(r^{\prime})}\, dr^{\prime\, 2} -
F^{\prime \, 2}(r^{\prime})\, d\phi^{\prime\, 2}
\end{equation}
with
\begin{equation}\label{30}
\begin{split}
t^{\prime\, 2} &= \frac{\bar{r}^2-Ml^2}{l^2}\, t^2\\
\\
r^{\prime\, 2} &= \frac{l^2}{\bar{r}^2-Ml^2}\, x^2\\
\\
\phi^{\prime\, 2} &= \frac{\bar{r}^2-Ml^2}{l^2}\, \phi^2
\end{split}
\end{equation}
where
\[x^2=r^2-\bar{r}^2\]
(the physical spacetime holds for $r\geq \bar{r}$ and $x\in [0,\infty]$)\\
and the value $r=\bar{r}$, for which $g_{\phi\phi}=0$, satisfies
the constraint
\begin{equation}\label{32a}
l\sqrt{M} < \bar{r} < l\sqrt{M+1}
\end{equation}
Also,
\begin{equation}\label{31}
F^{\prime\, 2}(r^{\prime})=r^{\prime\, 2}+\tilde{q}^2_m \ln
\left(\frac{r^{\prime\, 2}}{l^2}+1\right)
\end{equation}
\begin{equation}\label{32b}
\tilde{q}^2_m=Q^2_m e^{\bar{r}^2/Q^2_m}
\end{equation}
With $\tilde{q}_m=0$, the metric (\ref{29}) yields the AdS$_{3}$
background. However, this metric does not describe a magnetically
charged three-dimensional black hole, it is horizonless, it has no
curvature singularities, and it does not exhibit any signature
change. According to the interpretation of Cataldo et al \cite{catal}, a
two-dimensional solenoid carrying a steady current located at
spatial infinity can be considered as the source of the magnetic
field given by
\begin{equation}\label{33}
B(r)\sim \frac{1}{\sqrt{\frac{r^2}{l^2}+1}}
\hspace{1ex}.
\end{equation}
%%%%%%%%%%%%%%%%%%%%%%%%%%%%%%%%%%%%%%%%%%%%%%%%%%%%%%%%%%%%%%%%%%%%%%%%%%%%%%%%%%%%%%%%%%%%%%%%%
%%%%%%%%%%%%%%%%%%%%%%%%%%%%%%%%%%%%%%%%%%%%%%%%%%%%%%%%%%%%%%%%%%%%%%%%%%%%%%%%%%%%%%%%%%%%%%%%%
%%%%%%%%%%%%%%%%%%%%%%%%%%%%%%%%%%%%%%%%%%%%%%%%%%%%%%%%%%%%%%%%%%%%%%%%%%%%%%%%%%%%%%%%%%%%%%%%%
\section{Energy and Momentum Density Distributions}
The aim of this section is to evaluate the effective gravitational
mass of the radial magnetic field in a 2+1 Anti-de Sitter
spacetime (\ref{29}) using M{\o}ller's energy-momentum complex. We
first have to evaluate the superpotentials in the context of
M{\o}ller's prescription. There are four nonzero independent superpotentials
\bea\label{super}
\xi^{1\,2}_{1}&=&\frac{2}{l^2}\left(r^{2}+\tilde{q}^{2}_{m}\ln\left|1+\displaystyle{\frac{r^2}{l^2}}\right|\right)\nn\\
\xi^{2\, 1}_{1}&=&-\xi^{1\,
2}_{1}=-\frac{2}{l^2}\left(r^{2}+\tilde{q}^{2}_{m}\ln\left|1+\displaystyle{\frac{r^2}{l^2}}\right|\right)\\
\xi^{2\, 3}_{3}&=&-2\left(1+\frac{\tilde{q}^{2}_{m}}{l^2}+\frac{r^2}{l^{2}}\right)\nn\\
\xi^{3\,2}_{3}&=&-\xi^{2\,
3}_{3}=2\left(1+\frac{\tilde{q}^{2}_{m}}{l^2}+\frac{r^2}{l^{2}}\right).\nn
\eea

%\par\noindent
By substituting the M{\o}ller's superpotentials, as given by
(\ref{super}), into equation (\ref{mtheta}), one gets the energy
density distribution
\be \mathcal{J}^{0}_{0}=
\frac{r\left(1+\displaystyle{\frac{\tilde{q}^{2}_{m}}{l^2}}+\displaystyle{\frac{r^2}{l^{2}}}\right)}
{2\pi l^{2}\left(1+\displaystyle{\frac{r^2}{l^2}}\right)}
\label{energyden}
\ee
while the momentum density distributions
take the form \bea
\mathcal{J}^{0}_{1}&=&0\label{momden1}\\
\mathcal{J}^{0}_{2}&=&0\label{momden2}\hspace{1ex}. \eea

%\par\noindent
Therefore, if we substitute equation (\ref{energyden}) into
equation (\ref{menergy}), we get the energy of the radial magnetic
field in the 2+1 Anti-de Sitter spacetime that is contained in a
``sphere'' of radius $r_0$ \be E(r_0)=\frac{1}{4\pi
l^2}\left(r^{2}_{0}+\tilde{q}^{2}_{m}\ln\left|1+\frac{r^{2}_{0}}{l^2}\right|\right).
\label{energy} \ee This result is the effective gravitational mass
($E=M_{eff}$) of the spacetime under study.
%\par\noindent
Furthermore, if we independently evaluate the energy density
distribution for the AdS$_3$ metric using the M{\o}ller's
energy-momentum complex we get \be \mathcal{J}^{0}_{0}=
\frac{r}{2\pi l^2} \label{adsenergyden} \ee and by integration we
get the energy of the AdS$_3$ spacetime in a ``sphere'' of radius
$r_0$ \be E(r_0)=\frac{1}{4\pi l^2}r^{2}_{0} \hspace{1ex}.
\label{adsenergy} \ee It is evident that by setting
$\tilde{q}_{m}=0$ in equation (\ref{energy}) we get the energy of
the 2+1 Anti-de Sitter spacetime (see (\ref{adsenergy})) which was
independently evaluated  using M{\o}ller's prescription.
%%%%%%%%%%%%%%%%%%%%%%%%%%%%%%%%%%%%%%%%%%%%%%%%%%%%%%%%%%%%%%%%%%%%%%%%%%%%%%%%%%%%%%%%%%%%%%%%%
%%%%%%%%%%%%%%%%%%%%%%%%%%%%%%%%%%%%%%%%%%%%%%%%%%%%%%%%%%%%%%%%%%%%%%%%%%%%%%%%%%%%%%%%%%%%%%%%%
%%%%%%%%%%%%%%%%%%%%%%%%%%%%%%%%%%%%%%%%%%%%%%%%%%%%%%%%%%%%%%%%%%%%%%%%%%%%%%%%%%%%%%%%%%%%%%%%%
\section{Conclusions}
In this work, we have explicitly calculated the energy and
momentum densities associated with the magnetic solution to the ($2+1$) Einstein-Maxwell gravity.
The specific gravitational background describes the radial magnetic field in an
AdS$_3$ spacetime which is horizonless and it does not have any curvature singularities.
The magnetic solution depends on a ``charge'' $\tilde{q}_{m}$. By setting this ``charge''
to zero the magnetic solution to the ($2+1$) Einstein-Maxwell gravity becomes the
pure AdS$_3$ spacetime.
We employed M{\o}ller's prescription in order to compute the effective gravitational
mass, i.e. the total energy, contained in a ``sphere'' of radius $r_0$ in the
aforementioned gravitational background.
It should be stressed that the concept of effective gravitational mass is related
to the repulsive effects of gravitation.
Additionally, the corresponding momenta are zero due to the vanishing momentum density distributions.
Furthermore, we have independently computed the energy of the pure AdS$_3$ spacetime using again
 M{\o}ller's prescription. This result is identical to that obtained
when setting $\tilde{q}_{m}=0$ in the expression for the energy
of the magnetic solution to the ($2+1$) Einstein-Maxwell gravity.
\par
However, it should be pointed out that in the last years due to the AdS/CFT correspondence there has been much
progress in obtaining finite stress energy tensors of asymptotically AdS spacetimes\footnote{For a short review see
\cite{sken}.}. The gravitational stress energy tensor is in general infinite due to the infinite volume of the
spacetime. In order to find a meaningful definition of gravitational energy one should subtract the divergences.
The proposed prescriptions so far were ad hoc in the sense that one has to embed the boundary in some reference
spacetime. The important drawback of this method is that it is not always possible to find the suitable reference
spacetime. Skenderis and collaborators\footnote{Right after the first work of  Henningson and Skenderis
\cite{hen1}, Nojiri and Odintsov \cite{nojiri1} calculated a finite gravitational stress energy tensor for an
asymptotically AdS
spacetime where the dual conformal field theory is dilaton coupled. Furthermore, Nojiri and Odintson \cite{nojiri2},
and Ogushi \cite{nojiri3} found well-defined gravitational stress energy tensors for asymptotically AdS spacetimes in the
framework of higher derivative gravity and of gauged supergravity with single dilaton respectively. }
\cite{hen1,hen2,skenderis,haro}, and also Balasubramanian and Kraus \cite{bal}, described and implemented a new
method  which provides an intrinsic definition of the gravitational stress energy tensor. The computations are
universal in the sense that they apply to all asymptotically AdS spacetimes.
Therefore, it is nowadays right to state that the issue
of the gravitational stress energy tensor for any asymptotically AdS spacetime has been thoroughly understood.
\par\noindent
Finally, our results presented here provide evidence in support of Lessner's statement
\cite{Lessner} for the significance of M{\o}ller's prescription.
%%%%%%%%%%%%%%%%%%%%%%%%%%%%%%%%%%%%%%%%%%%%%%%%%%%%%%%%%%%%%%%%%%%%%%%%%%%%%%%%%%%%%%%%%%%%%%%%%%%%
%%%%%%%%%%%%%%%%%%%%%%%%%%%%%%%%%%%%%%%%%%%%%%%%%%%%%%%%%%%%%%%%%%%%%%%%%%%%%%%%%%%%%%%%%%%%%%%%%%%%
%%%%%%%%%%%%%%%%%%%%%%%%%%%%%%%%%%%%%%%%%%%%%%%%%%%%%%%%%%%%%%%%%%%%%%%%%%%%%%%%%%%%%%%%%%%%%%%%%%%%%%
%%%%%%%%%%%%%%%%%%%%%%%%%%%%%%%%%%%%%%%%%%%%%%%%%%%%%%%%%%%%%%%%%%%%%%%%%%%%%%%%%%%%%%%%%%%%%%%%%%%%%%
%%%%%%%%%%%%%%%%%%%%%%%%%%%%%%%%%%%%%%%%%%%%%%%%%%%%%%%%%%%%%%%%%%%%%%%%%%%%%%%%%%%%%%%%%%%%%%%%%%%%%
\section*{Acknowledgements}
The author is indebted to Dr. E.C. Vagenas for useful suggestions and comments.
This work was partially supported by Association EURATOM-Hellenic
Republic, University of Thessaly. The content of the publication is the sole responsibility of its
author and it does not necessarily represent the views of the Commission or its services.
%%%%%%%%%%%%%%%%%%%%%%%%%%%%%%%%%%%%%%%%%%%%%%%%%%%%%%%%%%%%%%%%%%%%%%%%%%%%%%%%%%%%%%%%%%%%%%%%%%%%%%
%%%%%%%%%%%%%%%%%%%%%%%%%%%%%%%%%%%%%%%%%%%%%%%%%%%%%%%%%%%%%%%%%%%%%%%%%%%%%%%%%%%%%%%%%%%%%%%%%%%%%%
%%%%%%%%%%%%%%%%%%%%%%%%%%%%%%%%%%%%%%%%%%%%%%%%%%%%%%%%%%%%%%%%%%%%%%%%%%%%%%%%%%%%%%%%%%%%%%%%%%%%%%
%%%%%%%%%%%%%%%%%%%%%%%%%%%%%%%%%%%%%%%%%%%%%%%%%%%%%%%%%%%%%%%%%%%%%%%%%%%%%%%%%%%%%%%%%%%%%%%%%%%%%%
%%%%%%%%%%%%%%%%%%%%%%%%%%%%%%% BIBLIOGRAPHY
%%%%%%%%%%%%%%%%%%%%%%%%%%%%%%%%%%%%%%%%%%%%%%%%%%%%%%%%%

%%%%%%%%%%%%%%%%%%%%%%%%%%%%%%%%%%%%%%%%%%%%%%%
%%%%%%%%%%%%%%%%%%%%%%%%%%%%%%%%%%%%%%%%%%%%%%%
%%%%%%%%%%%%%%%%%%%%%%%%%%%%%%%%%%%%%%%%%%%%%%%
\end{document}